\documentclass[pra,superscriptaddress,twocolumn,superscriptaddress,amsmath]{revtex4}
\usepackage{amsmath, amsthm, amssymb}
\usepackage[]{graphicx} 
\usepackage{caption}
\usepackage{dcolumn} 
\usepackage{bm} 
\usepackage[normalem]{ulem}
\usepackage{cases}
\usepackage{color}
\usepackage[dvipsnames]{xcolor}
\usepackage{algorithm}
\usepackage{algpseudocode}
\usepackage{csquotes}
\usepackage{mathtools}
\usepackage{mathrsfs}  
\usepackage{mathtools}
\usepackage[caption=false]{subfig}
\usepackage{lipsum}
\usepackage{comment}
\usepackage{tabularx}
\usepackage{multirow}
\usepackage{braket}
\usepackage{pgf} 
\usepackage{siunitx} 
\usepackage{caption}
\usepackage{physics}

\usepackage{hyperref}
\hypersetup{
    colorlinks=true,
    linkcolor=blue,
    citecolor=blue,
    filecolor=magenta,      
    urlcolor=cyan
    }

\usepackage[colorinlistoftodos]{todonotes}

\definecolor{pinocol}{rgb}{0,.4,1}

\begin{document}

\title{Countermeasures for Trojan-Horse Attacks on self-compensating all-fiber polarization modulator}

\author{Alberto De Toni}
\affiliation{Dipartimento di Ingegneria dell'Informazione, Università degli Studi di Padova, via Gradenigo 6B, IT-35131 Padova, Italy}

\author{Aynur Cemre Aka}
\affiliation{Dipartimento di Ingegneria dell'Informazione, Università degli Studi di Padova, via Gradenigo 6B, IT-35131 Padova, Italy}

\author{Costantino Agnesi}
\affiliation{Dipartimento di Ingegneria dell'Informazione, Università degli Studi di Padova, via Gradenigo 6B, IT-35131 Padova, Italy}

\author{Davide~Giacomo~Marangon}
\affiliation{Dipartimento di Ingegneria dell'Informazione, Università degli Studi di Padova, via Gradenigo 6B, IT-35131 Padova, Italy}

\author{Giuseppe Vallone}
\affiliation{Dipartimento di Ingegneria dell'Informazione, Università degli Studi di Padova, via Gradenigo 6B, IT-35131 Padova, Italy}
\affiliation{Padua Quantum Technologies Research Center, Università degli Studi di Padova, via Gradenigo 6A, IT-35131 Padova, Italy}

\author{Paolo Villoresi}
\affiliation{Dipartimento di Ingegneria dell'Informazione, Università degli Studi di Padova, via Gradenigo 6B, IT-35131 Padova, Italy}
\affiliation{Padua Quantum Technologies Research Center, Università degli Studi di Padova, via Gradenigo 6A, IT-35131 Padova, Italy}


\date{\today}

\begin{abstract}
Quantum Key Distribution (QKD) leverages the principles of quantum mechanics to exchange a secret key between two parties. Unlike classical cryptographic systems, the security of QKD is not reliant on computational assumptions but is instead rooted in the fundamental laws of physics. In a QKD protocol, any attempt by an eavesdropper to intercept the key is detectable: this provides an unprecedented level of security, making QKD an attractive solution for secure communication in an era increasingly threatened by the advent of quantum computers and their potential to break classical cryptographic systems.
However, QKD also faces several practical challenges such as transmission loss and noise in quantum channels, finite key size effects, and implementation flaws in QKD devices. Addressing these issues is crucial for the large-scale deployment of QKD and the realization of a global quantum internet. A whole body of research is dedicated to the hacking of the quantum states source, for example using \textit{Trojan-Horse attacks} (THAs), where the eavesdropper injects light into the system and analyzes the back-reflected signal. In this paper, we study the vulnerabilities against THAs of the \textit{iPOGNAC} encoder, first introduced in \cite{Avesani:iPOGNAC}, to propose adapted countermeasures that can mitigate such attacks.
\end{abstract}

\maketitle


\section{Introduction}
Quantum Key Distribution (QKD) represents one of the most mature applications of quantum information science, offering unconditional security for key exchange based on the fundamental laws of quantum mechanics \cite{bennettQuantumCryptographyPublic2014, Gisin2002}. Protocols such as BB84 and its decoy-state variants have been implemented over increasing distances and integrated into field-deployable systems \cite{lucamariniOvercomingRateDistance2018, Wang2022}. However, the practical security of QKD systems depends critically on the faithful implementation of their underlying quantum protocols. Any deviation or imperfection in components can introduce vulnerabilities that are not covered by theoretical security proofs \cite{curráslorenzo2025securityhighspeedquantumkey}.

One such vulnerability is posed by \textit{Trojan Horse Attacks} (THAs) \cite{Gisin2006, Jain2014}, in which an eavesdropper (Eve) injects bright light into the QKD apparatus, typically at the sender's (Alice’s) side, and then analyzes the back-reflected signal to gain information about secret settings, such as basis choices or intensity levels \cite{Vakhitov2001, Gisin2006, Jain2015}. These attacks exploit the physical layers of the system and, if undetected, can significantly compromise the security of the generated keys without introducing errors that are detectable by Alice and Bob (the receiver) \cite{qi2015}.

A growing body of research has focused on both theoretical models of THAs and practical countermeasures, including the use of optical isolators, monitoring detectors, and new security proofs that account for side-channel leakage \cite{Lucamarini2015, Sajeed2015, Gisin2006}. Despite these efforts, fully characterizing and mitigating THAs remains an open and crucial challenge in the secure deployment of QKD networks. 

In this paper, we analyze the practical impact of THAs on the \textit{iPOGNAC}, a self-compensating, all-fiber polarization modulation scheme, first proposed by Agnesi, Avesani et al. \cite{Avesani:iPOGNAC}. This analysis is of relevance since many different QKD systems have adopted the iPOGNAC since, both for \textit{Discrete-Variable} (DV) QKD, for polarization modulation \cite{Avesani2022} and time-bin modulation \cite{Scalcon2022}, and for \textit{Continuous-Variable} (CV) QKD \cite{sabatini2025}, showcasing the reliability, robustness and effectiveness of the polarization encoder. 
Other works previously analyzed the impact of THAs on this type of encoder, like \cite{Luo2024}, but without addressing the possibility of Eve directly exploiting the Sagnac-loop to gain information on the encoded symbol, a scenario in which she can get most of her optical power back.

In this work we perform an analysis of the attack in different regimes: continuous and pulsed laser, high and low mean photon number. We investigate the countermeasures that Alice can adopt in her system to drastically reduce the information leakage at Eve's side, give estimates of the limitations of such defenses and analyze the performances of the different types of THAs in those conditions.
\begin{figure*}
    \centering \includegraphics[width=0.55\linewidth]{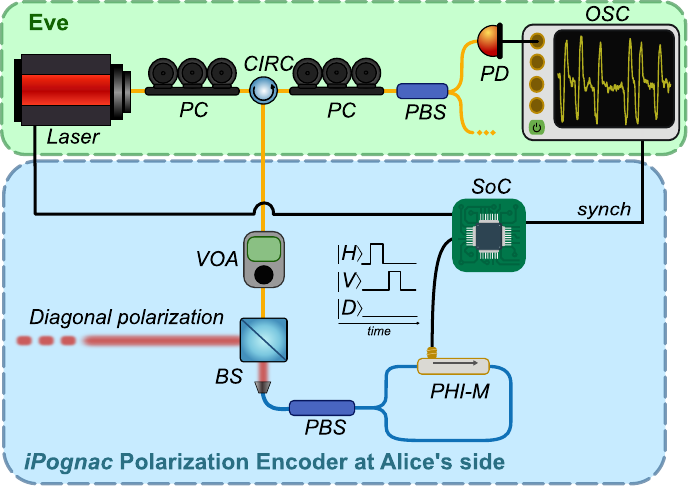}
    \caption{Scheme of the setup for the THA on the \textit{iPOGNAC}. SM-Fibers in yellow, PM-Fibers in blue, electrical connections in black. The optical power entering inside Alice's setup is indicated as $\mu_{in}$, while the one coming out (subject to the attenuations of the Alice's system and therefore smaller than $\mu_{in}$) is indicated as $\mu_{out}$.}
    \label{fig:setup}
\end{figure*}
\section{Methods}
Naïve implementation of Sagnac-loop based modulation schemes can be problematic since a significant portion of injected light through a THA can return to the eavesdropper containing an information leakage that can undermine the security of the implemented quantum communication protocol. In fact, the light injected by the eavesdropper traverses the same modulation loop Alice uses to encode her information, meaning that it potentially experiences the same modulation and carries the same information as the QKD signal, allowing the eavesdropper to gain knowledge on the forthcoming symbol. 

To analyze the performances of the THA on the iPOGNAC, we divided our work into three stages: an attack in strong-light regime, which comprises a continuous-wave laser attack (CWLA) and a pulsed laser attack (PLA) (both of them make use of photo-diodes as light detectors), a pulsed laser attack in weak-light regime exploiting Geiger-mode detectors such as \textit{Superconducting-Nanowire Single Photon Detectors} (SNSPDs) and an evaluation/analysis on the countermeasures that Alice can take to counteract these attacks. The reasons we chose to study these cases is because high-power attacks are simpler and cheaper to implement, whereas single-photon-level attacks allow for a wider range of countermeasure levels. In the meantime, pulsed attacks permit to concentrate more photons in a single time-span, while CW attacks require less stringent synchronization (more on the motivations for each attack in section \ref{sec:discussion}). The stages are reported in the following sections, after a brief description of the setup. 
\begin{figure*}
    \centering \includegraphics[width=1\linewidth]{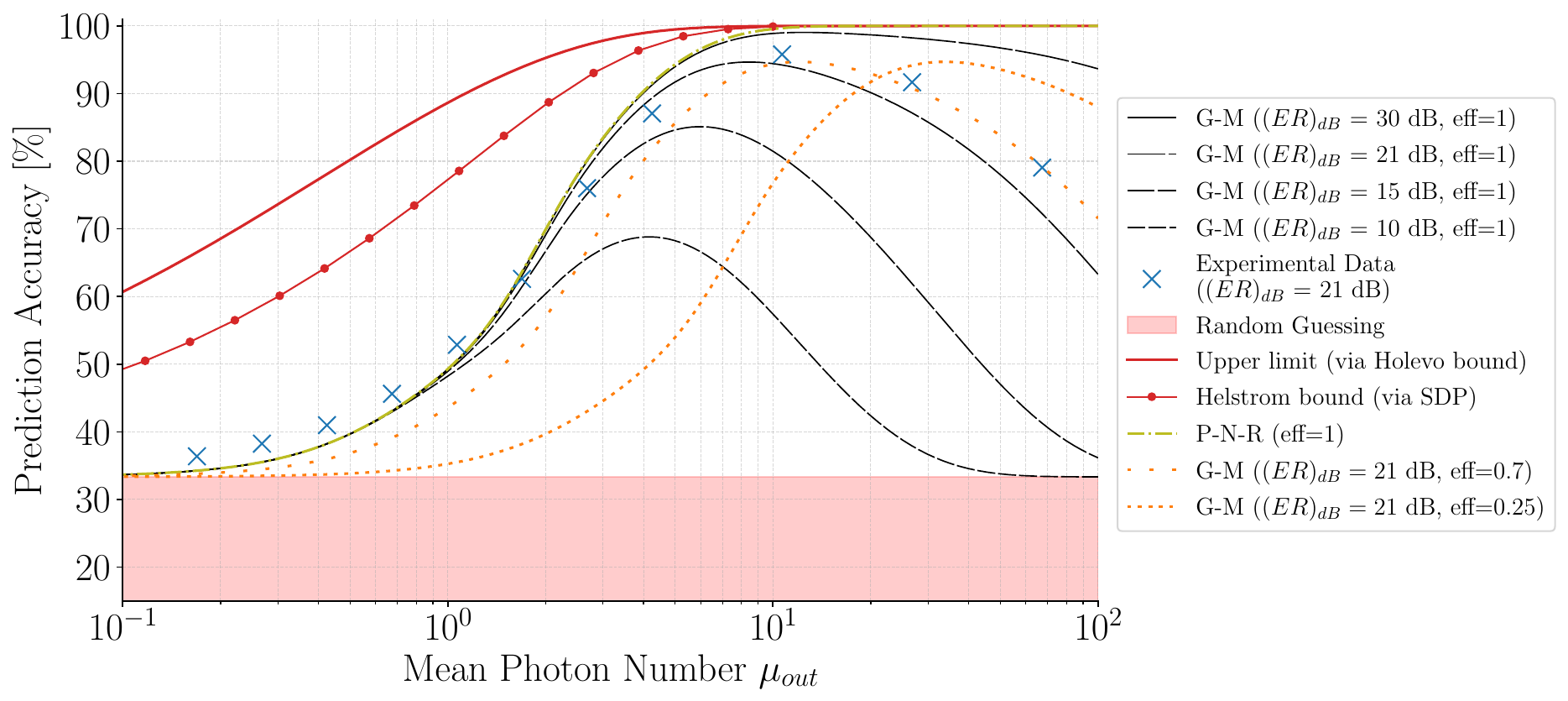}
    \caption{Theoretical prediction accuracy with respect to the mean photon number $\mu_{out}$, varying different POVMs, extinction ratios of the projective measurements, efficiencies, and experimental data for comparison. }
    \label{fig:predacc-mu-er}
\end{figure*}
\subsection{Description of the system}
The iPOGNAC polarization encoder is a device designed for stable, low-error, and calibration-free polarization modulation, particularly suited for QKD and quantum communications. Its core innovation lies in its use of a Sagnac interferometer loop, which inherently compensates for environmental disturbances such as temperature fluctuations and phase drifts, ensuring long-term operational stability.

Its working principle begins with linearly polarized optical pulses injected into the iPOGNAC. A half-wave plate rotates the polarization to a diagonal state, after which a beam splitter separates the input and output streams. The transmitted light is coupled into a polarization-maintaining fiber (PMF). The light then enters a fiber-based polarization beam splitter (PBS), which initiates the Sagnac loop. Here, the two orthogonal polarization components (horizontal and vertical) travel in opposite directions — clockwise (CW) and counter-clockwise (CCW) — along the slow axis of the PMF. Each component passes through a phase modulator at different times, allowing independent phase control. By applying specific phase shifts to the CW and CCW pulses, the iPOGNAC can generate all the polarization states required for the BB84 QKD protocol, including diagonal, anti-diagonal, and circular polarizations. Experimental results have demonstrated the iPOGNAC’s ability to maintain a low QBER over extended periods, both in laboratory and field-trial settings. The device’s versatility is further highlighted by its ability to be reconfigured for tasks such as time-bin encoding \cite{Scalcon2022}, continous variable modulation \cite{sabatini2025} and intensity modulation \cite{Berra2023}, supporting the development of robust and flexible quantum networks \cite{Avesani2021resource, Telebit}.
Regarding the attack, at Eve's side, a 1560 nm laser shines through a \textit{Polarization Controller} (PC) that controls the polarization in input at Alice's side. The light that comes back from Alice's system is redirected using a \textit{Circulator} (CIRC) to another PC and a \textit{Polarizing Beam Splitter} (PBS) that allows the projective measurement on a \textit{Si-Photodiode} (PD) or on the SNSPDs. The output of the PD is displayed using an \textit{Oscilloscope} (OSC), which is also used to store the data to analyze. Alice's setup is composed of a \textit{Variable Optical Attenuator} (VOA) to counteract the THA, and the asymmetric-iPOGNAC, namely a \textit{Beam Splitter} (BS) connected to Alice source on one side to a Sagnac-loop composed of a PBS and a \textit{Phase Modulator} (PHI-M) on the other side. Alice source emits
diagonal polarized photons that enter
the BS from the left side. For simplification purposes, the whole system (Alice+Eve) is controlled through the same \textit{System-On-A-Chip} (SoC), a \textit{Field-Programmable-Gate-Array} (FPGA) that allows to generate electrical pulses and synchronize everything to the same clock-rate \cite{monmasson2007fpga}: this holds on the generalized assumption that Eve has access to the system of Alice to exploit the same synchronization signals she uses, or that she can use clock-recovery algorithms to extract the information on the synchronization. The setup of the experiment is depicted in its entirety in Fig. \ref{fig:setup}. The modulation signals employed work by introducing a delay between the $\ket{H}$ and $\ket{V}$ symbols on the FPGA side: in this sense the Sagnac loop enables to encode the polarization states on the optical signal by fine-tuning this delay (see Fig. \ref{fig:hist2d}). We care to point out that since Alice is using an attenuator as countermeasure, the light of the THA coming out from her system will experience two times the attenuation set on the VOA. In our setup, the total attenuation experienced by the light coming out from the system $\text{Att}_{tot}$ can be described by this equation:
\begin{equation}
    (\text{Att}_{tot})_{\text{dB}} = 2(\text{Att}_{VOA}+\Delta A) + 6 + E
\end{equation}
where $\text{Att}_{VOA}$ is the attenuation manually set on the VOA (the term we mostly refer to in this paper when talking about countermeasures), $\Delta A$ is an inefficiency term given by the characterization of the VOA (in our case is 4 dB), $6$ dB is the attenuation given by the double crossing of the 50:50 beam splitter, and $E$ are extra losses from coupling inefficiencies, defective components, and so on, in our case corresponding to $\sim 1$ dB.


\section{Theoretical Model}
\label{sec:theoretical_model}

This section is dedicated to the calculation of the guessing probability in function of the output mean photon-number $\mu_{out}$. We will cover three cases: using an optimal POVM for the measurement, which gives an upper-bound on the maximum information that can be extracted from the system, while using a fixed POVM, photon-number resolving detectors in the second case and Geiger-mode photodetectors in the third case. All of them are reported in the following sections. Despite the theoretical model being valid both for weak and strong light regimes, the main limitations of the latter mostly come from the limited performances of the devices in use, like the noise floor of the detectors or the oscilloscopes (see sec. \ref{sec:resultsstronglight} and \ref{sec:discussion}).

\subsubsection{Optimal POVM}
When implementing a THA, Eve will manage to obtain a state that depends on the symbol modulated by Alice. With no lack of generality, we therefore assume that the states that Eve will receive are 
\begin{equation}
\label{eq:psij}
    \begin{aligned}
    \ket{\psi_1}&=\ket{\sqrt{\mu}}_H\otimes\ket{0}_V \quad &&\text{if Alice sends $\ket{H}$,} \\
    \ket{\psi_2}&=\ket{0}_H\otimes
    \ket{\sqrt{\mu}}_V \quad 
    &&\text{if Alice sends $\ket{V}$,} \\
    \ket{\psi_3}&=
    \ket{\sqrt{\tfrac\mu2}}_H\otimes \ket{\sqrt{\tfrac\mu2}}_V \quad &&\text{if Alice sends $\ket{D}$.}
\end{aligned}
\end{equation}
Optimizing on the POVM $F_e$ employed by Eve, it's possible to extract an upper bound on the quantity of mutual information that is accessible and can be extracted from the system, the so-called \textit{accessible information}:
\begin{equation}
\label{eq:Iacc}
    I_{\rm acc}(A:E) = \max_{\{F_e\}}\sum_{a,e}p(a,e)\log_2\dfrac{p(a,e)}{p(a)p(e)}
\end{equation}
with $p(e|a)={\rm Tr}[F_e\rho_a]$ and $\rho_a$ is the density matrix of the received state $\ket{\psi_a}$. Instead, the average probability of correctly guessing a symbol $p_{g}$, in this scenario, is given by:
\begin{equation}
p_g=\sum_a p(a)p(e=a|a)
\end{equation}
Assuming the system is symmetric,
namely $p(a)=1/3$, $p(e=a|a)$ and $p(e\neq a|a)$ do not depend on $a$, then $p(e)=1/3$ and:
\begin{equation}
\label{eq:helstrom}
    p^*_g={\rm Tr}[F^*_a\rho_a] \quad \forall a
\end{equation}
where $F^*_a$ is the optimal POVM. Therefore the accessible information can be related to the guessing probability by:
\begin{equation}
I_{\rm acc}(A:E) =p_g\log_2 (3p_g)+(1-p_g)\log_2 \frac{3(1-p_g)}{2}
\label{eq:Iacc2}
\end{equation}
While the maximization in  \eqref{eq:Iacc} is in general hard to solve because there is not an explicit closed form for the optimal POVM, the accessible information can be upper-bounded by the so called Holevo bound \cite{Weedbrook2012}, 
which expresses the amount of classical information that can be extracted from a quantum system, obtained as a mixture $\rho=\sum_ap_a\rho_a$:
\begin{equation}
\label{eq:Holevo_bound}
\begin{aligned}
    I_{\rm acc}(A:E) \leq &\chi(\rho) 
    =S(\rho)-\sum_ap_aS(\rho_a)
    \\
    =S(\rho)
    \end{aligned}
\end{equation}
The last equality follows from the fact the state $\rho$ is given by $\rho=\frac13\sum_{a=1}^3\ketbra{\psi_a}{\psi_a}$ and the Von Neumann entropy of a pure state is vanishing (i.e.: $S(\ketbra{\psi_a}{\psi_a})=0$).
Then, it follows that eq. (\ref{eq:Iacc}) is upper bounded by $S(\rho)$.


Solving this equation can yield an upper bound for the $p_g$. As a first step we therefore have to calculate $S(\rho)$, that is equal to the Shannon Entropy
of the eigenvalues $\lambda_i$ of the matrix given by $\rho_{ij}=\frac{1}{3}\braket{\psi_i}{\psi_j}$
(see appendix \ref{appendix}):

\begin{align}
    S(\rho)=H(\mu):=-\sum_{i=1}^3 \lambda_i(\mu)\log_2(\lambda_i(\mu))
\end{align}
By using eq. \eqref{eq:Holevo_bound}, an upper bound on the guessing probability $p_g$ can be extracted in function of $\mu$ by the implicit relation:
\begin{align}
    I_{acc}(A:E)\leq H(\mu)
\end{align}
which gives the results reported in fig. \ref{fig:predacc-mu-er} as a red line.

The results just found depict a loose upper bound on the actual, tighter, quantum limit for the distinguishability of a set of states, which is given by the Helstrom bound \cite{Helstrom1969QuantumDA, Weedbrook2012} (originally defined for bi-partite systems), which quantifies the maximum probability $p^*_g$ of identifying the correct state. 
To compute this limit for the states identified by the density matrices $\{\rho_i\}$, it is possible to express the primal-form optimization problem:
\begin{equation}
\begin{aligned}
    p_g^*&=\frac13\max_{F_a}\sum_a {\rm Tr}[F^*_a\rho_a]  \\
    F_a \succeq &0 \quad \forall\ a \quad \text{and} \quad \sum_aF_a = \mathbb{I}
\end{aligned}
\end{equation}
This semi-definite program (SDP) is strictly feasible \cite{Tavakoli2024} (for instance, the trivial measurement $F_a=\mathbb{I}/3$ satisfies the constraints) which means that the Slater condition for convex optimization is satisfied \cite{BenTal2001}, granting that Strong Duality holds and we can write this as a dual-problem SDP \cite{Assalini2010}. This method allows us to optimize over all generalized quantum measurements to find the optimal strategy by finding a Hermitian operator $K$ that solves the following minimization problem (in general computationally simpler to solve): 

\begin{equation}
p_g = \min_K\Big\{\text{Tr}(K)|K\succeq p_i \rho_i, \quad \forall i\Big\} \quad 
\end{equation}
where $p_i=\frac{1}{3}$ in our case.

The solution to this program allows to relax the bound on $\mu$ with which an eavesdropper can get useful information on the system, and is reported in Fig. \ref{fig:predacc-mu-er} as a red line with dots.



\subsubsection{Fixed POVM}
For this attack, we wanted to minimize the amount of detectors whilst still being able to distinguish the symbols, which corresponds to a minimum of two channels, as seen from the truth-table of the detections reported in Table \ref{tab:truthtable}:

\begin{table}[H]
\centering
\begin{tabular}{c c|c c}
   & \textbf{detector}\hspace{0.2em} &  $CH_1$  &  $CH_2$  \\
   \hspace{1em}\textbf{input}\hspace{1em} & & & \\ 
   \hline
    $\ket{H}$ & & \hspace{10pt}click\hspace{10pt} & \hspace{10pt}no-click\hspace{10pt} \\
    $\ket{V}$ & & \hspace{10pt}no-click\hspace{10pt} & \hspace{10pt}click\hspace{10pt} \\
    $\ket{D}$ & & \hspace{10pt}click\hspace{10pt}  &  \hspace{10pt}click\hspace{10pt}
\end{tabular}
    \caption{Truth-table of the detections (clicks) of the symbols in a Geiger-mode detector using two channels.}
    \label{tab:truthtable}
\end{table}

Based on the table, we can already predict that the minimum amount of photons per symbol required to distinguish all the three symbols is two (therefore the two channels, to be able to distinguish $\ket{D}$). Knowing the probability of a channel to \textit{not click} is described by a Poissonian distribution \Big($\mathcal{P}_\mu(0)=\dfrac{\mu^0}{0!}e^{-\mu}=e^{-\mu}$\Big), we can describe the guessing probabilities for each symbol in the following way:

\begin{equation}
\begin{split}
    P\big(\ket{H}_{out}\ |\ \ket{H}_{in}\big) &= P(CH_1\ \cap\ \overline{CH_2}) \\ 
    &\stackrel{i.i.d.}{=} P(CH_1)\cdot P(\overline{CH_2}) \\ 
    &= \big(1-e^{-\mu_H\eta}\big)\cdot e^{-\mu_H\eta\ \cdot ER} \\ \\
\end{split}
\end{equation}
and similarly for $\ket{V}$ and $\ket{D}$:
\begin{equation}
\begin{split}
    P\big(\ket{V}_{out}\ |\ \ket{V}_{in}\big) 
    & = e^{-\mu_V\eta \cdot ER}\cdot \big(1-e^{-\mu_V\eta\ }\big) \\ \\
    P\big(\ket{D}_{out}\ |\ \ket{D}_{in}\big) 
    &= \big(1-e^{-\mu_D\eta}\big)^2 \\
\end{split}
\end{equation}
where $\eta$ is the efficiency of the detectors and $\mu_H$, $\mu_V$ and $\mu_D$ are, respectively, the mean photon numbers when sending $\ket{H}$ and projecting the measurement on $CH_1$, sending $\ket{V}$ and projecting on $CH_2$, and sending $\ket{D}$ - projecting detector should be irrelevant. Assuming the detectors are equally balanced, i.e. $\mu=\mu_H=\mu_V=2\mu_D$, and the efficiency is maximum, i.e. $\eta=1$, the equations can be written as:

\begin{equation}
    \begin{split}
        P\big(\ket{H}_{out}\ |\ \ket{H}_{in}\big) &= P\big(\ket{V}_{out}\ |\ \ket{V}_{in}\big) \\ 
        &= \big(1-e^{-\mu}\big)\cdot e^{-\mu\cdot ER} \\ \\
        P\big(\ket{D}_{out}\ |\ \ket{D}_{in}\big) &= \big(1-e^{-\mu/2}\big)^2
    \end{split}
\end{equation}

\begin{figure}[h] 
    \centering \includegraphics[width=1.0\linewidth]{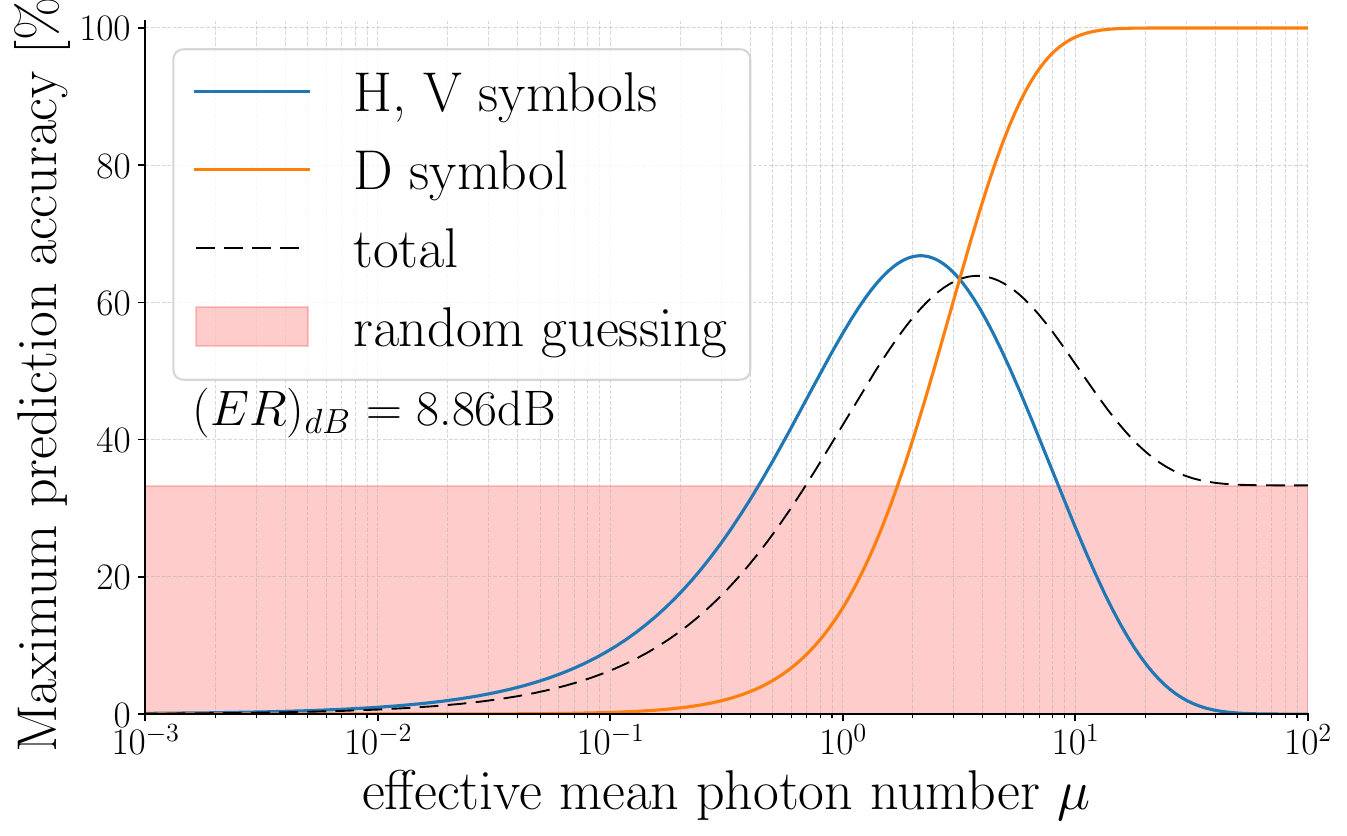}
    \caption{Theoretical detection probabilities for each state and the total with respect to the mean photon number $\mu_{out}$ in the use-case of only two detectors (Extinction Ratio = 8.86 dB).}
    \label{fig:predacc-mu}
\end{figure}

Summing the three cases, we obtain the total detection probability we can expect from a certain extinction ratio (i.e.: prediction accuracy). The dependency of the prediction accuracy from mean photon number $\mu$ and extinction ratio $ER$ (defined in linear scale as the ratio of $|\alpha|^2$ and $|\beta|^2$ in $\ket\psi=\alpha\ket H+\beta\ket V$, and $ER = 10^{-(ER)_{dB}/10}$) can be graphically visualized in Fig. \ref{fig:predacc-mu} and Fig. \ref{fig:predacc-mu-er}.

Let's suppose Eve is in possession of a perfect photon-number-resolving detector with optimal detection efficiency. We can model Eve's approach by describing the back-reflected photons as coherent states of the form:
\begin{equation}
    \ket{\Psi_E}=\dfrac{1}{\sqrt{2}}\big(\ket{\mu_{out}^H}+e^{i\varphi}\ket{\mu_{out}^V}\big)
\end{equation}
Where $\mu_{out}$ is the mean photon number coming back from Alice's system and $\varphi$ holds the information on the encoding state given by the phase modulator.
At $\mu_{out}=0$ Eve is forced to randomly guess the symbol, therefore her $P^E_{guess}(0)=1/3$. Already from $\mu_{out}=1$, it can be seen that for Eve is more convenient to use two channels for detection, because her $P^E_{guess}(1)=2/3$, given that if she detects a photon on $\ket{H}$ or $\ket{V}$ her best strategy is to keep the received state, as she will guess correctly two times out of three. Even though she knows that the state she received could also be a $\ket{D}$ state, a random guess between these two symbols will yield a $P^E_{guess}=\frac{1}{2}<\frac{2}{3}$. Intuitively, increasing the number of detectors will just cause an increase in the uncertainty. For $\mu_{out}\ge2$, we fall into the case described later in Table \ref{tab:truthtable}, therefore her $P^E_{guess}(\ge2)$ will depend on the detection probabilities for each channel, namely $\dfrac13\left(\Pr(H|H)+\Pr(V|V)+\Pr(D|D)\right)$. In tab. \ref{tab:detprobs} we listed the detection probabilities $\Pr(\text{Eve}|\text{Alice})$ both for Photon-Number-Resolving and imperfect Geiger-Mode photodetectors, noting as $c(\nu)=(1 - e^{-\nu})$ and $\bar{c}(\nu)=e^{-\nu}$ the functions, respectively, indicating the probability for a detector to click and not-click.
\begin{table}[h]
\centering
\begin{tabular}{c | c | c}
    Probability & G-M & P-N-R \\
    \hline
    \hline
    $\Pr(H|H)$ & $c(\mu_{out})\cdot \bar{c}(\mu_{out} ER)$ & $c(\mu_{out})$ \\
    $\Pr(V|H)$ & $c(\mu_{out}ER)\cdot \bar{c}(\mu_{out})$ & $0$ \\
    $\Pr(D|H)$ & $c(\mu_{out})\cdot c(\mu_{out}ER)$ & $0$ \\
    $\Pr(\text{vac}|H)$ & $\bar{c}(\mu_{out})\cdot\bar{c}(\mu_{out}ER)$ & $\bar{c}(\mu_{out})$ \\
    & & \\
    $\Pr(H|V)$ & $=\Pr(V|H)$ & $0$ \\
    $\Pr(V|V)$ & $=\Pr(H|H)$ & $c(\mu_{out})$ \\
    $\Pr(D|V)$ & $=\Pr(D|H)$ & $0$ \\
    $\Pr(\text{vac}|V)$ & $=\Pr(\text{vac}|H)$ & $\bar{c}(\mu_{out})$ \\
    & & \\
    $\Pr(H|D)$ & $c(\mu_{out}/2)\cdot \bar{c}(\mu_{out}/2)$ & $c(\mu_{out}/2)\cdot \bar{c}(\mu_{out}/2)$ \\
    $\Pr(V|D)$ & $=\Pr(H|D)$ & $=\Pr(H|D)$\\
    $\Pr(D|D)$ & $c(\mu_{out}/2)\cdot c(\mu_{out}/2)$ & $c(\mu_{out}/2)\cdot c(\mu_{out}/2)$ \\
    $\Pr(\text{vac}|D)$ & $\bar{c}(\mu_{out}/2)\cdot\bar{c}(\mu_{out}/2)$ & $\bar{c}(\mu_{out}/2)\cdot\bar{c}(\mu_{out}/2)$
\end{tabular}
\caption{Detection probabilities for each (Eve$|$Alice) case using Geiger-Mode (G-M) and Photon-Number-Resolving (P-N-R) photodetectors. We considered a more realistic scenario for G-M detectors, introducing also a sub-optimal extinction ratio ($ER$, in linear scale) coming from the projecting device at Eve's side. We identified as ``vac" the vacuum state, where no detector clicks.}
\label{tab:detprobs}
\end{table}
Given the aforementioned probabilities that Eve correctly guesses a symbol at a certain $\mu_{out}$, we can derive the calculation of the total $P_{\text{guess}}^E(\mu_{out})$ as follows:
\begin{equation}
\begin{split}
    &P_{\text{guess}}^E(\mu_{out})= \\
    &=\dfrac{1}{3}\Pr(\langle0|\mu_{out}\rangle)+\dfrac23\Pr(\langle1|\mu_{out}\rangle)+\\
    &\hspace{1.2em}+P^E_{guess}(\ge2)\cdot\left(1-\Pr(\langle0|\mu_{out}\rangle)-\Pr(\langle1|\mu_{out}\rangle)\right)\\
    &=\frac{1}{3} e^{-\mu_{out}(\mu_{out} +1)} \Big[\mu_{out} ^2+\\
    &\hspace{1.2em}+e^{\mu_{out} ^2} \left(-2 e^{\mu_{out} /2}+3 e^{\mu_{out} }-1\right)+\\
    &\hspace{1.2em}+2 e^{\mu_{out} /2} (\mu_{out} ^2+1\Big)-e^{\mu_{out} } \left(\mu_{out} ^2+2\right)+1\Big]
\end{split}
\end{equation}
The resulting guessing probability $P^E_{guess}$ is displayed for the ideal photon-number-resolving detector with optimal extinction ratio as a green dot-dashed line, and for the Geiger-mode detector with different detection efficiencies and extinction ratios, in Fig. \ref{fig:predacc-mu-er}.



\section{Strong light regime}
The attack in strong light regime has been performed employing commercially-available photo-diodes. These not only have bandwidths that allow for a full resolution of the spectral response, but can also estimate proportionally the quantity of incoming light, admitting a wide range of attacks, contrarily from the weak light regime where Geiger-mode-type photo-detectors must be employed, limiting the possibilities. We acknowledge the fact that the methods hereby presented are sub-optimal, and don't use all the information available from the total shape of the outputs. Machine-Learning classifiers can be adopted to improve the results. For the scope of this paper, more centered on the countermeasures to these types of attacks, we will leave the introduction of Machine-Learning methods as a task for future works.
\subsection{CW laser attack} 
\begin{figure}[h]
    \centering \includegraphics[width=0.95\linewidth]{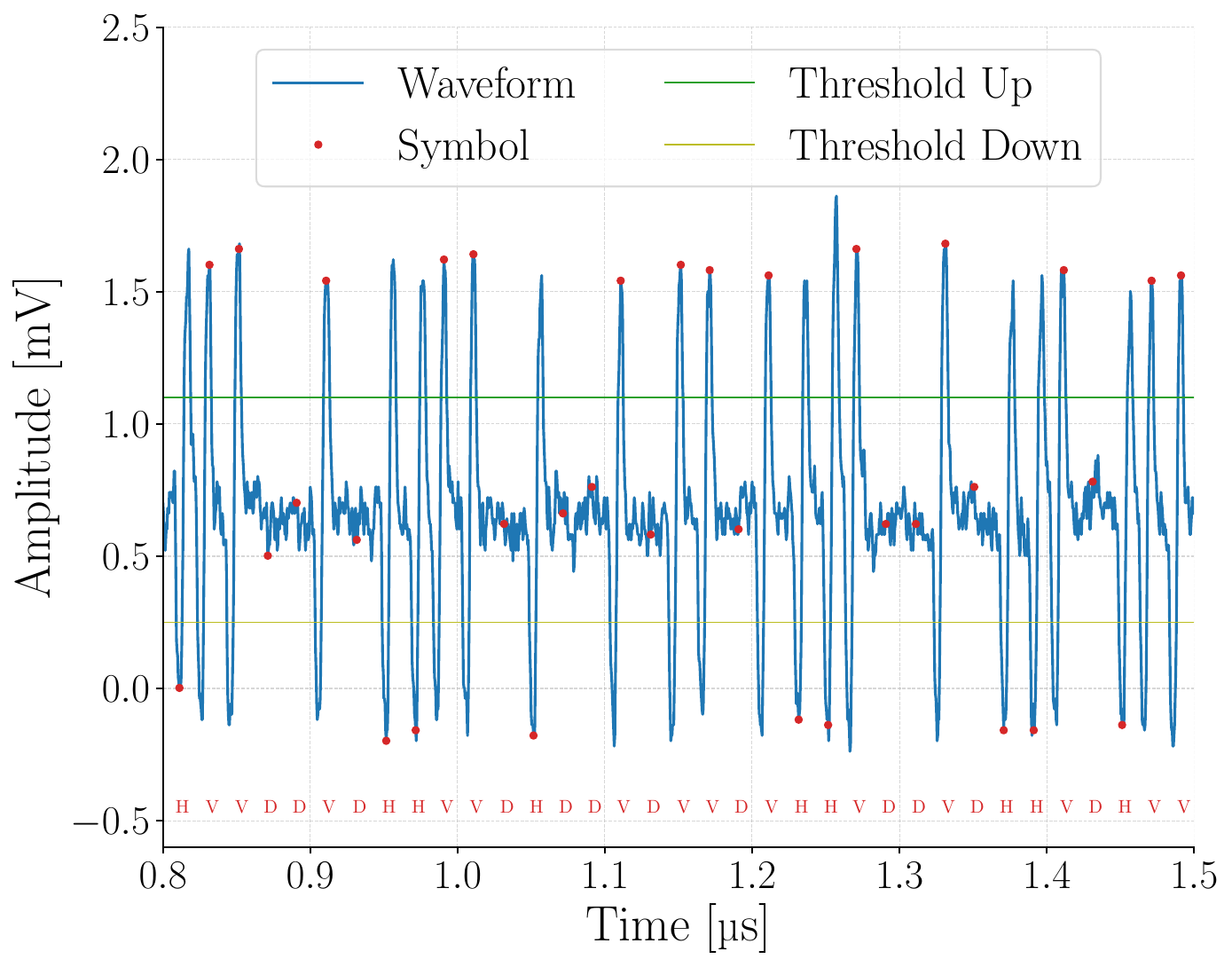}
    \caption{Waveform with calculated symbol positions and classified symbols for the CWLA. Red dots show the estimated symbol locations, while thresholds are used to classify the symbol as $\ket{H}$, $\ket{V}$, or $\ket{D}$ (if the symbol location is above the threshold up, the symbol is classified as $\ket{V}$, if it is between the thresholds up and down, it is classified as $\ket{D}$, and if it is below the threshold down, it is classified as $\ket{H}$).}
    \label{fig:seq_reconstruction_cont}
\end{figure}

For the hacking in continuous regime, we made sure to fine-tune the first PC to enter Alice's setup with the $\ket{D}=\frac{1}{\sqrt{2}}\big(\ket{H}+\ket{V}\big)$ state. The second, analogously, serves to project the $\ket{D}$ state on the $\ket{H}$ and $\ket{V}$ state by means of the PBS and measure them with the photodiode. The resulting waveform should be equally split above and below the average, as it is visible on the screen of the oscilloscope reported in Fig. \ref{fig:seq_reconstruction_cont}. 

\begin{figure}[h]
    \centering \includegraphics[width=0.95\linewidth]{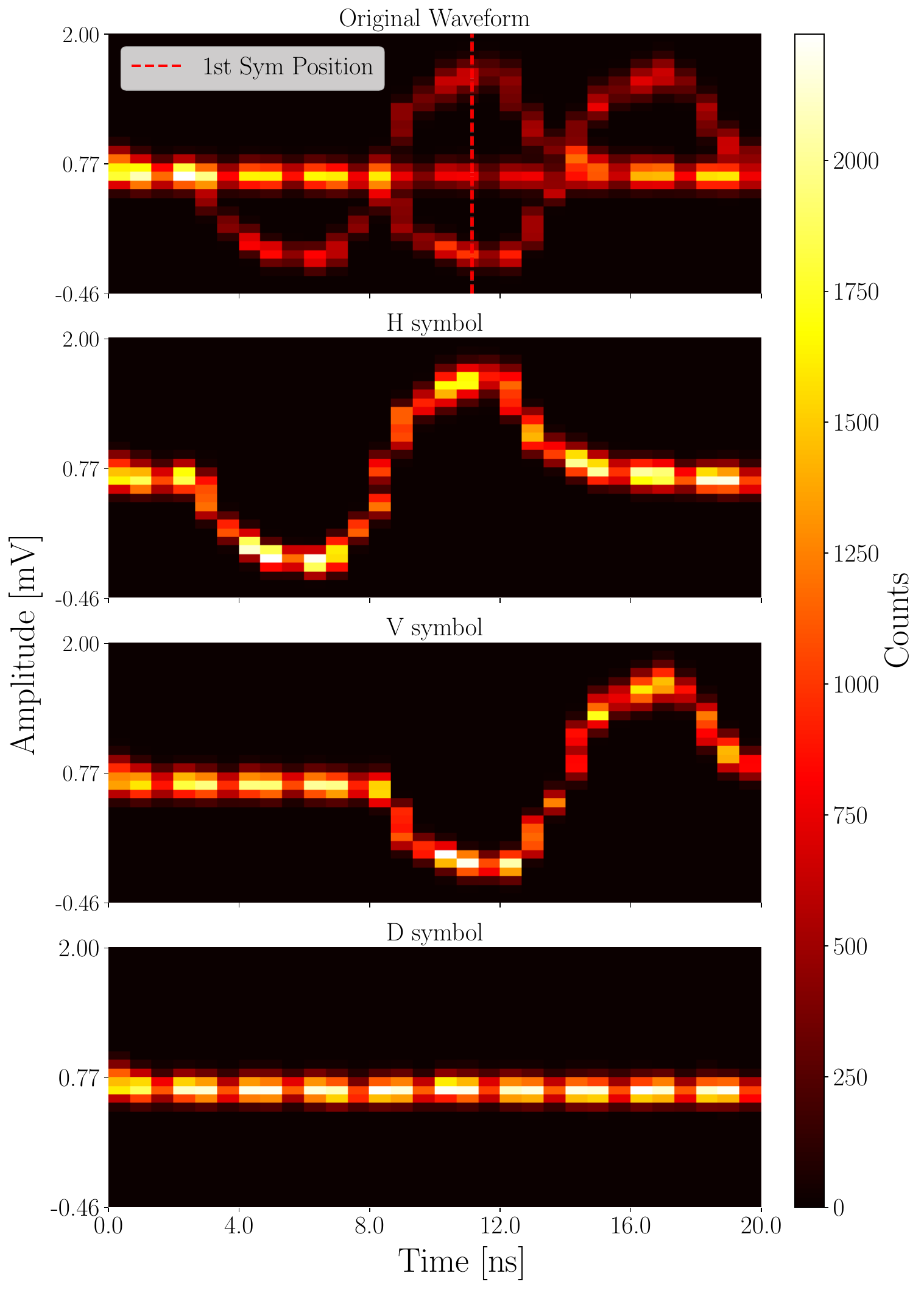}
    \caption{2D-histogram of the waveform resulting from the oscilloscope, result of the optical modulation in continuous wavefront by the \textit{iPOGNAC} during the THA, and $\ket{H}$, $\ket{V}$ and $\ket{D}$ symbols separately. All the waveforms presented are calculated by performing a modulo-period operation. In the top plot, the red vertical line shows the detected position of the first symbol in the original waveform.}
    \label{fig:hist2d}
\end{figure}

By taking that same plot and performing a ``modulo-period" operation (in this case the period being $20$ ns, since the repetition rate is $50$ MHz), we obtain a superposition of the $\ket{H}$ and $\ket{V}$ symbols in the exact location in time where the symbols are (in the timespan of one period) in the original waveform. We perform a 2D-histogram of these data, as for higher attenuations the shapes of the waveform get lost in the noise, obtaining a waveform characterized by the previously-mentioned down-up behavior, like the one depicted in Fig. \ref{fig:hist2d}. From there we can easily estimate the location of the first symbol in the sequence, which automatically gives us all the following symbol locations by simply increasingly adding the period (20 $ns$).

After estimating all symbol locations, each symbol is classified as $\ket{H}$, $\ket{V}$, or $\ket{D}$ using thresholds, as shown in Fig. \ref{fig:seq_reconstruction_cont}.
The thresholds are chosen modeling the arrival of the heights of the symbols as Gaussian distributions with certain mean $\nu$ and variance $\sigma^2$ and using the \textit{Bayes decision rule for minimum error}, which guarantees the lowest possible probability of error under known distributions and priors \cite{fukunaga2000introduction} and yields that the optimal threshold \( t^* \) minimizing the total error is:
\[
t^* = \arg \min_{t} E(t)
\]
where $E(t)$ is the error function. 
In this sense, the best threshold happens to be found in the intersection point between the two Normal distributions, because that's where the likelihood of observing the value $x$ is the same for both the PDFs. In our scenario, the distributions of the $\ket{H}$, $\ket{V}$ and $\ket{D}$ symbols have very similar variances $\sigma_1^2\simeq\sigma_2^2=\sigma^2$. Therefore, the optimal threshold $t^*$ can be obtained as the middle point of their averages $\nu_i$:
\[
\frac{1}{\sigma \sqrt{2\pi}} \exp\left( -\frac{(t^* - \nu_1)^2}{2\sigma^2} \right) = 
\frac{1}{\sigma \sqrt{2\pi}} \exp\left( -\frac{(t^* - \nu_2)^2}{2\sigma^2} \right)
\]
Therefore $t^* = \dfrac{\nu_1 + \nu_2}{2}$.

\subsection{Pulsed laser attack} 
\begin{figure}[h]
    \centering \includegraphics[width=0.95\linewidth]{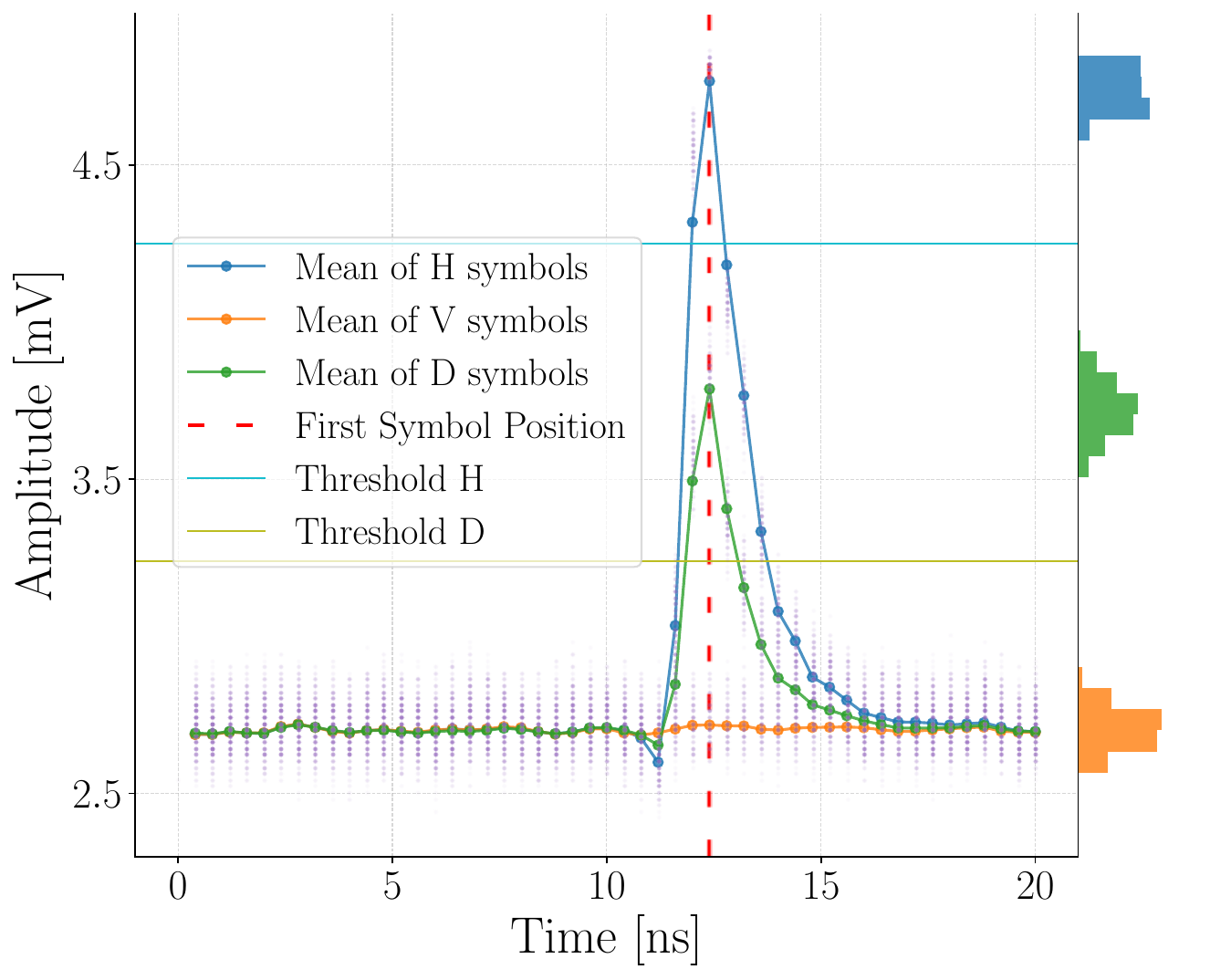}
    \caption{Mean of $\ket{H}$, $\ket{V}$, and $\ket{D}$ symbols in time modulo period, result of the optical modulation in pulsed regime by the \textit{iPOGNAC} during the THA. Data points are reported in purple. Red vertical line shows the detected position of the first symbol in the original waveform\iffalse: along the line, a 0\% intensity is associated to a $\ket{V}$ state, 50\% to a $\ket{D}$ state, 100\% to an $\ket{H}$ state\fi. Statistical distributions of these symbols are displayed on the right.}
    \label{fig:first-symbol-pulsed}
\end{figure}

For hacking in pulsed regime, following a method similar to that used in the previous section, a “modulo-period” operation is applied to the waveform, resulting in a superposition of the $\ket{H}$, $\ket{V}$ and $\ket{D}$ symbols, as illustrated in Fig. \ref{fig:first-symbol-pulsed} (inside the pulse the three heights are visible, $\ket{V}$ at $0\%$, $\ket{D}$ at $50\%$ and $\ket{H}$ at $100\%$). Once the position of the first symbol is estimated by locating the maximum, all symbol positions are determined by iteratively adding the period.

\begin{figure}[h]
    \centering \includegraphics[width=0.95\linewidth]{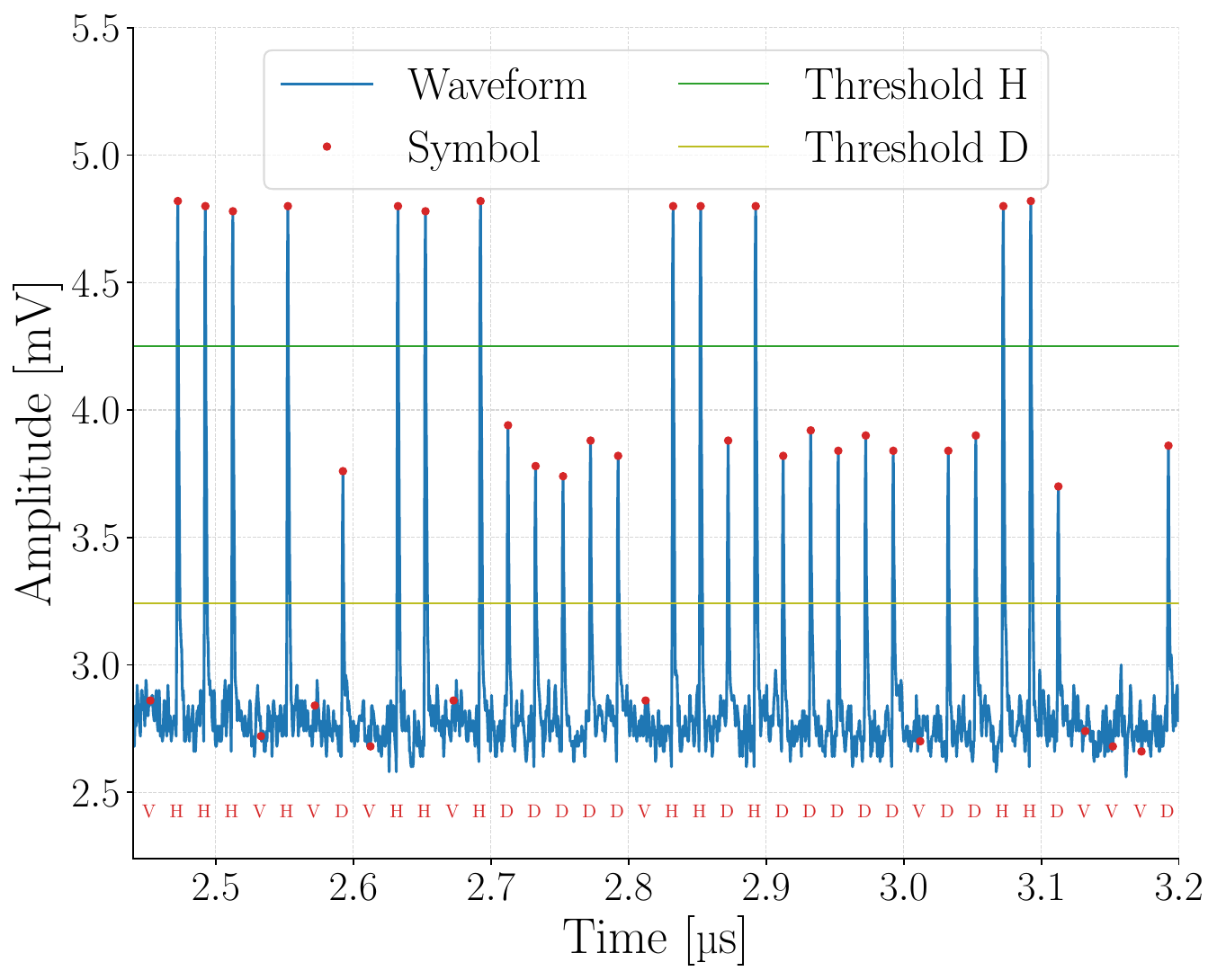}
    \caption{Waveform with calculated symbol positions and classified symbols for the pulsed laser attack. Red dots show estimated symbol locations, thresholds are used to classify the symbol as $\ket{H}$, $\ket{V}$, or $\ket{D}$ (if the symbol is above the threshold H, the symbol is classified as $\ket{H}$, between the thresholds H and D, it is classified as $\ket{D}$, otherwise it is classified as $\ket{V}$).}
    \label{fig:seq-reconstruction-p}
\end{figure}

As shown in Fig. \ref{fig:seq-reconstruction-p}, each symbol can be classified as $\ket{H}$, $\ket{V}$ or $\ket{D}$ using thresholds, after all the symbol locations have been determined. Thresholds are chosen using Gaussian distributions of the symbols and Bayes decision rule for minimum
error, mentioned in the previous section.

\subsection{Results for the Strong Light Regime}
\label{sec:resultsstronglight}
The result of optimal prediction accuracy for continuous and pulsed laser attacks is shown in Fig. \ref{fig:accuracy-attenuation}: 
\begin{figure}[H]
    \centering \includegraphics[width=1.0\linewidth]{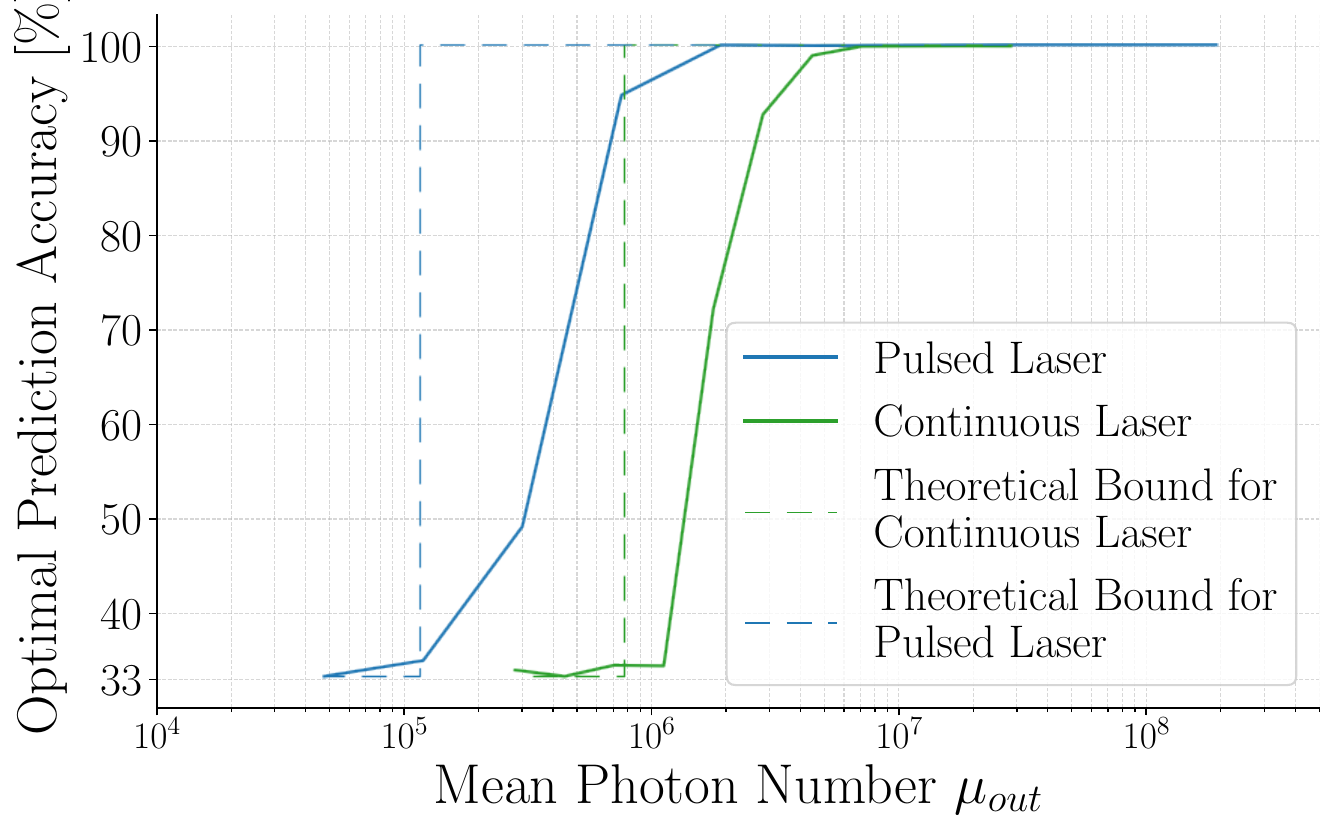}
    \caption{Results on optimal prediction accuracy based on the mean photon number $\mu_{out}$ for the continuous and pulsed laser attack.}
    \label{fig:accuracy-attenuation}
\end{figure}
For the CW laser, we observed that the sequence can be constructed from the theoretical bound of $\mu_{\text{out}} \sim 10^{7}$, which corresponds up to 8 dBs of attenuation.

From the setup scheme shown in the Fig. \ref{fig:setup}, the laser shines from Eve's side passes through the VOA twice. Consequently, the effective attenuation on Eve's laser is doubled. Notably, for this test we used an EDFA to increase the optical signal power in the case of the pulsed laser attack, which led to a noticeably higher average power of 60 mW. When compared to the CW case, this amplification allows for a gain of roughly $\sim$16.5 dB in the pulsed scenario, which increases Eve's capacity to compromise the transmission.

\begin{figure}[h]
    \centering \includegraphics[width=1.0\linewidth]{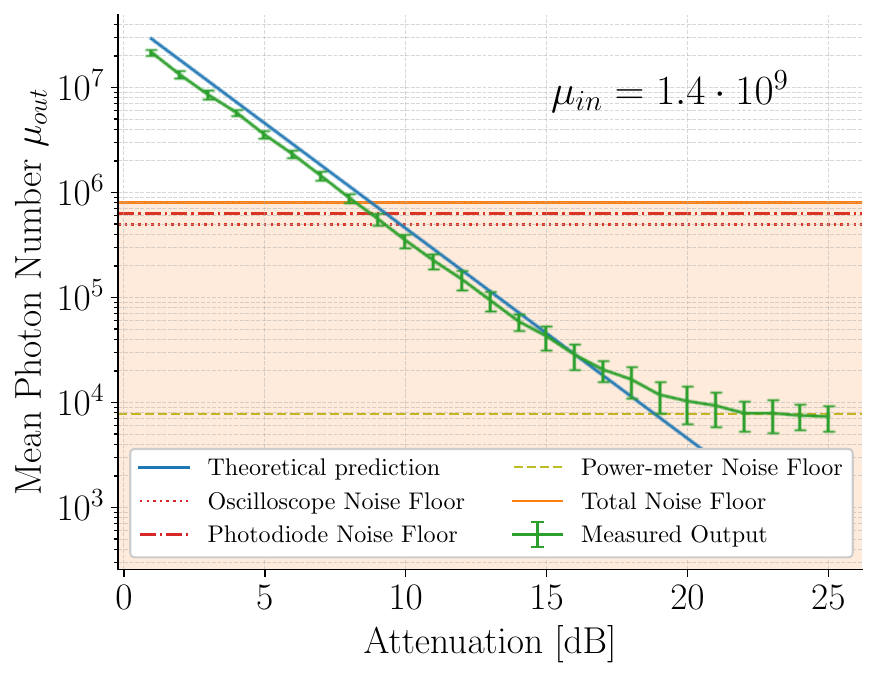}
    \caption{Mean photon number $\mu_{out}$ at different attenuations set on the VOA. Noise floors thresholds are indicated with horizontal lines. The plateau after 15 dB is due to the noise floor of the power-meter we used to characterized the output power (50 nW).}
    \label{fig:output-power}
\end{figure}

The mean photon number variation based on the attenuation set on the VOA is shown in Fig. \ref{fig:output-power}.
The theoretical output is calculated on the basis of the losses that occur in our setup. The total noise floor is calculated by the \textit {Root Sum Square} (RSS) of the oscilloscope and photodiode standard deviations, namely $\sigma_{\text{osc}}$ and $\sigma_{\text{pd}}$ as follows:
\begin{equation}
\sigma_{\text{total}} = \sqrt{\sigma_{\text{osc}}^2 + \sigma_{\text{pd}}^2} \simeq 5\ \mu W
\label{eq:totalnoisefloor}
\end{equation}

As seen in Fig. \ref{fig:output-power}, around 8 dBs, the output power of the CW laser is below the total noise floor, implying that the sequence cannot be predicted after this bound, consistent with the theoretical bound for the continuous laser shown in Fig. \ref{fig:accuracy-attenuation}. 

\section{Weak light regime}
In Geiger-mode photodetectors like SNSPDs, the duration of time required for the current to fully return to the nanowire or for the voltage to drop to the baseline noise level is known as the detector's dead (reset) time \cite{autebert2020}. 
The detectors we used have a dead time \(T_{dead}\) of $\sim 20$ ns, therefore the maximum achievable repetition rate \( f_{\text{max}} \) for the pulsed signal can be: 

\begin{equation}
    f_{\text{max}} \approx \frac{1}{T_{\text{dead}}} \approx 50\ \text{MHz}
    \label{eq:maxreprate}
\end{equation}
Given that during the period of dead time the detectors are unable to detect the incoming photons, there is a strong upper bound on the maximum achievable repetition rate, that is not only given by Eq. \ref{eq:maxreprate}, but also by the saturation of the SPDs. In fact, a repetition rate of $50$ MHz would yield meaningful results (i.e.: at least two photons per pulse) only in a situation of maximum saturation of the single channels of the SNSPDs. Since this could lead to irreparable damages to the equipment, we opted to lower the repetition rate to 1 MHz to account for the dead time of the SNSPDs and lower the expected counts.

An alternative to counteract this problem is to exploit multiple channels of the SNSPDs and a $1\times N$ beam splitter per PBS output. This could lead to an increase in the count rate, at the cost of increasing the number of channels used and at a required temporal alignment at the beginning, hypothetically doubling the maximum frequency with $N$. Since our analysis wants to tackle the issue with the less components as possible, we won't delve into this possibility, leaving it for future analyses. 
\subsection{Results for the Weak Light Regime}
Regarding the THA using SNSPDs as detectors, we measured the extinction ratio $(ER)_{dB}$ of the PBS at our disposal (measuring $\sim21$ dB), and predicted the sequence for different levels of attenuation. In Fig. \ref{fig:predacc-mu-er}, not only the experimental results are visible, but also how different $ER$s highly influence the trend of the theoretical prediction accuracy curve in variation of $\mu$. Our results show that a maximum of $\sim$95\% prediction accuracy was achieved in near-perfect conditions of $\mu$.

\begin{figure}[h]
\centering \includegraphics[width=1.0\linewidth]{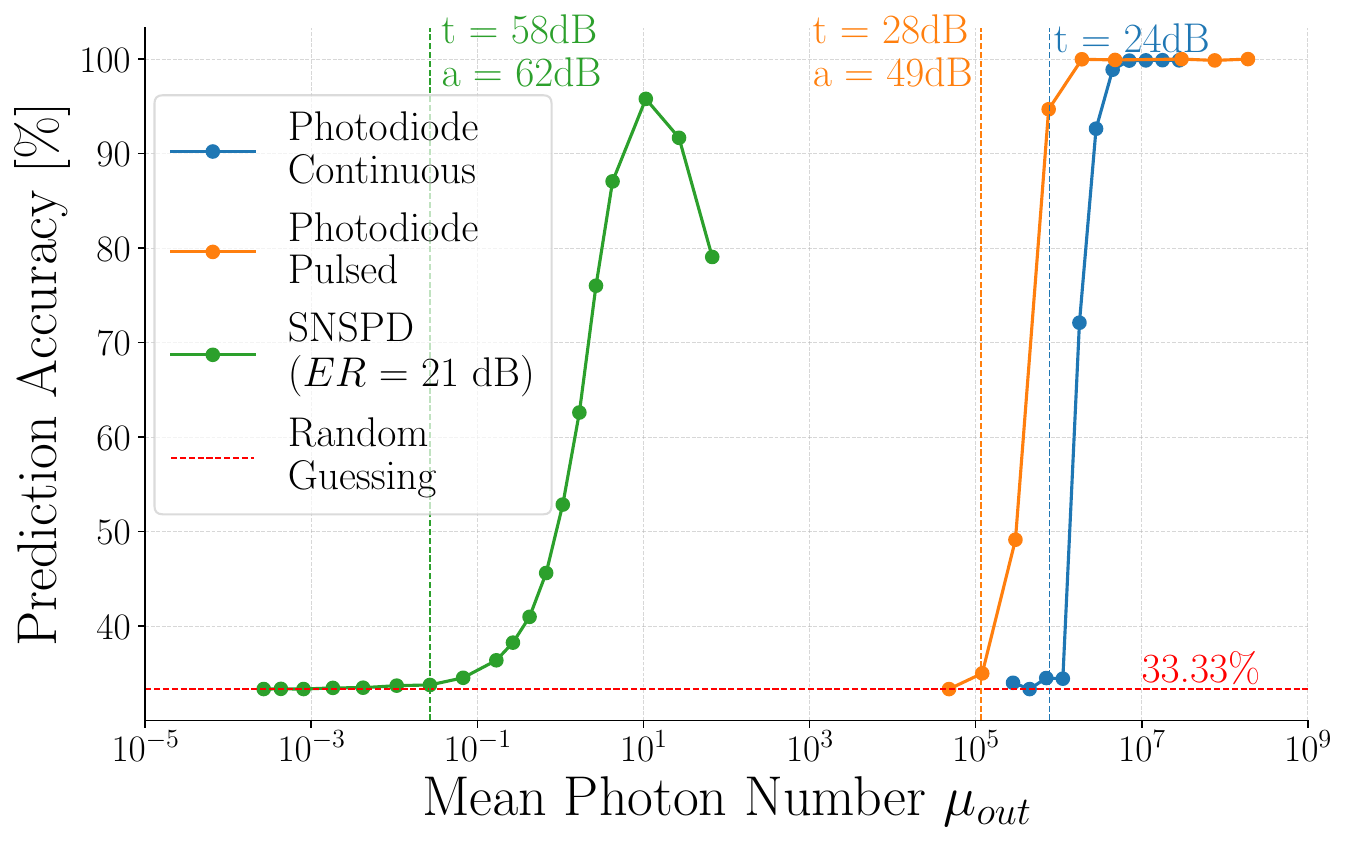}
    \caption{Prediction accuracy in relation to the mean photon number $\mu_{out}$, after normalizing the average power to 10 W for thermal damages ($t$) and 1 MW for ablation damages ($a$), for which the respective countermeasure attenuation is reported at the top.}
    \label{fig:predacc-att-norm}
\end{figure}

All these tests were performed in normal conditions of average power to give an estimate of the prediction accuracies. Potentially, real-world attackers can increase the power of their laser up to the maximum possible ratings of the components at their disposal (e.g.: the maximum power a fiber can handle, which normally is around 10 W for continuous regime 
and 1 MW for pulsed regime \cite{peterka2021high, smith2009}). For this purpose, we normalized the average powers of the results to a higher value to show the attenuation levels that are required by Alice to cope with such intensities. The results are collected and shown in Fig. \ref{fig:predacc-att-norm}, where it shows that an attenuation of $\sim$60 dB involved in a THA won't lend meaningful results to Eve, even when the eavesdropper is employing single-photon detectors in their setup.
\section{Comparison of the attacks}
\label{sec:discussion}
We studied a comparison of the attacks proposed in this paper, and collected them based on their complexities in Table \ref{table:comparision}. 

\begin{table}[h]
\centering
\begin{tabular}{|c c|c c c|}
\hline
   & \textbf{complexity}\hspace{0.2em} &  \textit{low}  &  \textit{medium} & \textit{high}  \\
   \hspace{0.2em}\textbf{regime}\hspace{0.2em} & & & & \\ 
   \hline
   \hline
    \textit{strong} & & PD-CWLA\hspace{1em} & PD-PLA\hspace{0.5em} & \\
    \textit{weak} & & & & G-M \\
    \hline
\end{tabular}
    \caption{Use-cases of the attacks proposed in this paper: Photodiode - continuous wave laser attack (PD-CWLA), Photodiode - pulsed laser attack (PD-PLA) and attack using Geiger-mode single-photon detectors (G-M).}
\label{table:comparision}
\end{table}
The attack in continuous-wave doesn't require the alignment of the pulses with Alice's setup, differently from the pulsed-laser one. Once Alice and Eve are synchronized, achieved either having an electrical connection to Alice's SoC or exploiting clock-recovery algorithms like \cite{calderaroFastSimpleQubitBased2020}, the continuous stream of photons provides all the information on the symbols being transmitted. Yet, this attack is the least resistant to countermeasures, given that a considerable amount of optical power is wasted in useless information between two subsequent symbols. On the other hand, the pulsed-regime attack requires precise alignment of Eve's pulses with Alice's modulation at the beginning of the transmission. Once that is achieved, though, the stream is straightforward and the use of photo-diodes (that can have bandwidths in the order of 2-20 GHz) guarantees every symbol of the sequence can be identified. It is important to acknowledge that all the results of the strong light regime are dependent on the total noise floor shown in Eq. \ref{eq:totalnoisefloor} and Fig. \ref{fig:output-power}, sum of the ones of the photo-diode and the oscilloscope. Improving the performances of these two components, it is possible to counteract more attenuation and reconstruct the sequence with better accuracies in strong light regime. As shown in Fig. \ref{fig:predacc-att-norm}, we might conclude that the pulsed-laser attack yields meaningful results for just some dBs of attenuation more than the continuous-wave attack. In reality, achieving higher peak powers is much easier than achieving higher average powers, for example using an \textit{Erbium-Doped Fiber Amplifier} (EDFA), or by reducing the duty-cycle of the pulses. Another advantage is that using a pulsed laser one can work closer to the thermal upper limit imposed by silica SM fibers, which is 1 MW for pulsed lasers \cite{smith2009, peterka2021high}.  
In comparison to the photo-diode used in the strong light scenario, attacks using SNSPDs are notably more complex, as reported in Table \ref{table:comparision}. The main reason for this complexity is the interaction of many parameters that control the reliability and efficiency of SNSPDs. Moreover, key performance indicators such as the extinction ratio after the states are measured (which dependancy is shown in Fig. \ref{fig:predacc-mu-er}), dark count rate, jitter time, reset time, etc. must be carefully optimized in tandem to ensure accurate and efficient single-photon detection \cite{tripathy2024snspd}. Better results on the prediction accuracy, using the same POVM, can be obtained by exploiting photon-number-resolving single-photon detectors, as we discussed in sec. \ref{sec:theoretical_model}.
\section{Countermeasures}
\label{sec:countermeasures}
Up to now we only considered passive countermeasures that involved attenuation at the entrance of Alice's setup using a VOA. This is because we investigated different types of THA counterattacks and categorized the most common defenses in:

\begin{itemize}
    \itemsep-0.2em
    \item \textbf{Passive}
        \begin{enumerate}
            \itemsep-0.2em
            \item Filtering (using a \textit{Wavelength Division Multiplexer} filter in Alice's transmission band);
            \item Isolation (using an Isolator or a Circulator at Alice's output);
            \item Attenuation (using Optical Attenuators at Alice's output);
        \end{enumerate}
    \item \textbf{Active}
        \begin{enumerate}
            \itemsep-0.2em
            \item \textit{Watchdog} detectors (exploiting a Circulator connected to a detector at Alice's output).
        \end{enumerate}
\end{itemize}

Passive countermeasures can be all traced back to the attenuation type, since all of them can be overcome sending enough optical power \cite{Ponosova2022, Vakhitov2001}. By measuring input and output power on different off-the-shelf components, we calculated for passive filtering and isolation to correspond to an approximate $\sim$60 dB of optical attenuation. The results that we reported confirm that even using SNSPDs as a mean of attack is not enough to get useful results on a real-world-scenario QKD setup, since the transmission rates are generally higher (transmission frequencies higher than 50-100MHz are sufficient with state-of-the-art equipments) and an isolator is often sufficient to block the incoming light in Alice's system. Furthermore, reducing back-reflections is a proactive measure that enhances overall security. This can be achieved by using angle-polished connectors (FC/APC) instead of flat connectors (FC/PC), eliminating open ports, and fusing connections where possible. These methods can limit attack opportunities regardless of the spectral characteristics of individual components \cite{Jain2015}.

An active countermeasure system can also be used, exploiting what are usually known as ``\textit{Watchdog}" detectors. These are devices that measure the light impinging in Alice's setup by means of a Circulator situated at the output of the system. These are particularly helpful because an attacker can be immediately spotted by measuring its optical power (that is generally high). There are three cases a Watchdog detector (WD) and an eavesdropper detector (ED) can face:

\begin{enumerate}
    \itemsep-0.2em
    \item \textit{The noise floor of the WD is lower than the one of the ED}: in this case the attacker is caught and the communication terminates;
    \item \textit{The noise floor of the WD is higher than the one of the ED}: in this case Alice could not perceive the presence of the attacker and can be hacked;
    \item \textit{The noise floors of the WD and of the ED are comparable}: in this case the attacker is surely spotted, since it needs to shoot much higher power to cope with Alice's setup internal attenuations.
\end{enumerate}
Therefore, an eavesdropper should aim at using detectors with very low noise level (such as single-photon detectors) to counteract the use of Watchdog detectors. Despite that, the effectiveness of both monitoring detectors (as well as the other passive countermeasures) can be limited by their spectral responsiveness, potentially allowing attacks at wavelengths where their sensitivity is negligible \cite{Jain2015}. In this sense, having Eve sending short pulses could limit the visibility of the Watchdog detector, not triggering the power measure. In conclusion, a near-optimal countermeasure to a THA can be achieved by mixing all of the proposed above.

\begin{figure}[h]
    \centering \includegraphics[width=1.0\linewidth]{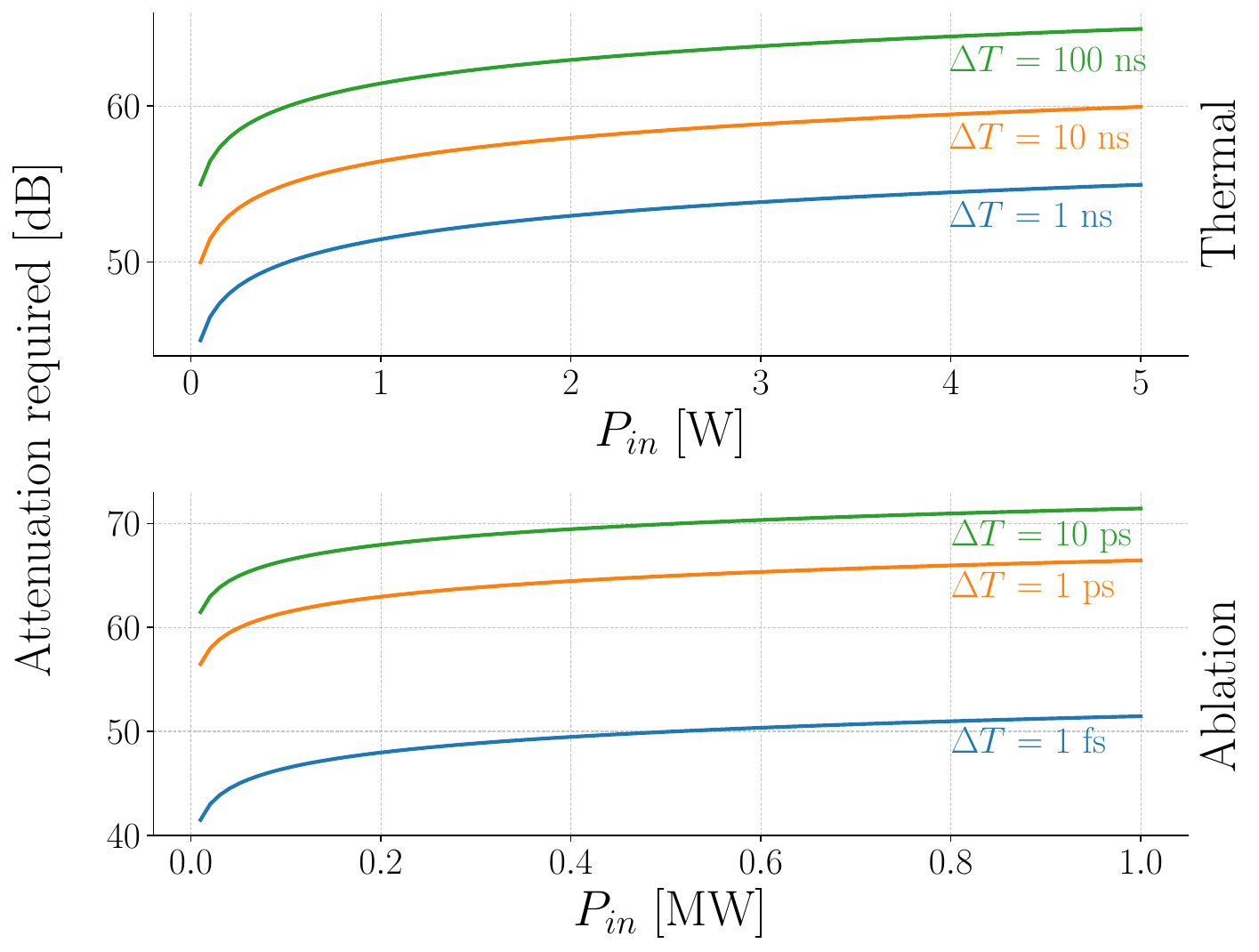}
    \caption{Attenuation required to counteract the THA presented in this work in different scenarios of input peak power $P_{in}$ and pulse width $\Delta T$, both for thermal (top) and ablation (bottom) damage limits.}
    \label{fig:countermeasures}
\end{figure}

We now try to estimate a lower bound on the attenuation required to counteract these attacks based on the assumption that the \textit{Laser-Induced Damage Threshold} (LIDT) of the system is set at 10 W for thermal damages and 1 MW for ablation damages. As shown in Fig. \ref{fig:predacc-mu-er}, a mean photon number of $\mu_{out}\sim5-8$ photons per symbol is sufficient for Eve to obtain a successful attack with a prediction accuracy higher than $80\%$ in good conditions of extinction ratio. In order to have full security on the \textit{iPOGNAC}, a stricter bound can be placed at $\mu_{out}<0.1$ photons per symbol to get Eve's prediction accuracy closer to the random guessing limit of 33\%. Therefore, we can give an estimate on the countermeasures that can be taken in these conditions. Assuming a total loss inside the iPOGNAC (excluded the VOA) in the best case (not accounting coupling losses or imperfections in the components) of $\Delta P\simeq6\ dB$, we can derive the amount of attenuation at the output that can counteract the attack:
\begin{equation}
    A_{\text{dB}} = \dfrac{1}{2}(10\log_{10}\left(\dfrac{\mu_{in}}{\mu_{out}}\right)-\Delta P)
\end{equation}
Where the $1/2$ is because of the optical pulse crossing two times the attenuation imposed by the VOA, and $\mu_{in}$ is solely dependent by parameters of the attacker:
\begin{equation}
    \mu_{in} = \dfrac{\lambda P_{in}\Delta T}{hc}
\end{equation}
where $h$ is Planck's constant (6.6261E-34 J$\cdot$s), $c$ is the speed of light in vacuum (2.9979E8 m/s), while $P_{in}$, $\lambda$ and $\Delta T$ are respectively the optical peak power in watts, the wavelength and the width of the pulse of Eve's laser in seconds. In Fig. \ref{fig:countermeasures} the required levels of attenuation required to counteract different levels of $P_{in}$ are shown. With $\sim65-70$ dB of attenuation, i.e. $\sim130-140$ dB of isolation, at the output, we can consider the iPOGNAC to be secure against the types of THA presented in this paper, which is in line with what presented by Lucamarini et al. in \cite{Lucamarini2015}. On average, the output power coming out of current QKD systems employing the \textit{iPOGNAC} revolves around $\mu\simeq10^5$ photons per symbol, therefore already requiring an attenuation of around $50$ dB to reach a typical mean photon number of 0.6 photons per symbol. As we have shown with our work, this is already a reasonable amount of countermeasure that deprives Eve of a lot of information. In order to be completely secure from this types of attacks, adding an isolator at the output (which, when traveled in the opposite direction adds an attenuation of at least $30$ dB) can be a good habit for state-of-the-art QKD systems in general.  

\section{Conclusion}
We analyzed different methodologies to approach Trojan-Horse Attacks that can be executed on a QKD setup that exploits three-state efficient-BB84 and a modulation scheme like the one of the \textit{iPOGNAC}, a self-compensating all-fiber polarization encoder first proposed in \cite{Avesani:iPOGNAC}. We proposed two different categories of attack, in strong and and weak light regime, where in the former the detection can be done exploiting photo-diodes while in the latter higher-efficiency single-photon detectors must be used. Even though the attack types are different in the two regimes (utilizing a more deterministic approach in one and a more probabilistic one depending on the mean-photon-number in the other) we estimated that the weak regime can be considered successful with countermeasures that comprise attenuations $\sim30\ dB$ higher than the ones required in the strong light regime, allowing to detect the sequence even in conditions of extremely low average power. However, the prediction accuracy is not as stable and deterministic as in the strong regime because of the detector types (the ones we adopted are Si-Photodiodes). Using photodiodes, in fact, the prediction accuracy follows a shape similar to an ``inversed sigmoid function", from 100\% to 33.33\% (equivalent to a random guessing), while for single-photon detectors the prediction accuracy is highly limited by the saturation of the detectors and the extinction ratio of the polarization states.

We've shown that an effective and immediate Trojan Horse Attack is possible in continuous waveform regime (without having to adjust the delays between a light pulse and the modulation of Alice), that the same kind of attack in pulsed regime can overcome more easily higher defense layers, and that attacks are also possible at higher attenuation levels using single-photon detectors with high sensitivity like SNSPDs. We investigated different types of countermeasures that can be applied to the setup that can limit the leaked information at Eve's side with ease. In a broader sense, we've estimated that having more than $\sim 65-70$ dB of attenuation ($\sim130-140$ dB of isolation) at Alice's output is many cases a good habit to limit the effects of Trojan Horse Attacks like the ones proposed, and that this can be improved in combination with wavelength filtering or optical isolation. We gave an estimate of the theoretical mutual information that can be extracted by the attack in weak regime, both in an hypothetical scenario using an optimal POVM, using perfect photon-number-resolving detectors and in a more realistic scenario using SNSPDs. For future works, we are planning the implementation of a Machine-Learning classifier to distinguish the states in strong light regime. 

\appendix

\appendix
\section{}
\label{appendix}

In this section we show how
to compute the
Von Neumann entropy of the state $\rho$
received by Eve, where
$\rho=\frac13\sum_{j=1}^3\ket{\psi_j}\bra{\psi_j}$
and the states $\ket{\psi_j}$
are defined in eq. \eqref{eq:psij}.
The Von Neumann entropy of $\rho$
is equal to the Shannon entropy of its eigenvalues.
Therefore, we need to evaluate the eigenvalues of
$\rho$.

Since the three states $\{\ket{\psi_1}, \ket{\psi_2}, \ket{\psi_3}\}$
are linear independent they can be considered as the first three elements of the basis.
Since the Hilbert space is infinite dimensional, we can complete the basis with orthonormal vectors $\{\ket{\psi_4} \dots \ket{\psi_\infty}\}$ that are orthogonal to the subspace defined by $\ket{\psi_j}$ with $(j=1,2,3)$.




Since $\rho$ has support only in the three-dimensional
subspace span by $\{\ket{\psi_j}\}$, we can restrict our attention to such subspace, in which the basis elements
$\rho_{jk}$
of $\rho$ are given by the following equation:
\begin{equation}
\label{eq:rhoInPsi}  
{\rho}\ket{\psi_k}=\sum_j \rho_{j k}\ket{\psi_j }
\end{equation}


 since 
$\rho=\frac13\sum_{j=1}^3\ket{\psi_j}\bra{\psi_j}$ we have that:
$$
\begin{aligned}
\rho\ket{\psi_k}&=\frac13\sum_{j=1}^3\ket{\psi_j}\braket{\psi_j}{\psi_k}
=
\sum_{j=1}^3\left(\frac13\braket{\psi_j}{\psi_k}\right)\ket{\psi_j}
\end{aligned}
$$
Comparing with \eqref{eq:rhoInPsi}  it is clear that:
$$
\rho_{jk}=\frac13\braket{\psi_j}{\psi_k}
$$
Therefore, the matrix $\rho$ in the $\{\ket{\psi_j}\}$ basis is given by:
\begin{align}
    \rho = \dfrac13\begin{pmatrix}
        1 & e^{-\mu} & e^{-\mu(1-\frac{1}{\sqrt{2}})} \\
        e^{-\mu} & 1 & e^{-\mu(1-\frac{1}{\sqrt{2}})} \\
        e^{-\mu(1-\frac{1}{\sqrt{2}})} & 
        e^{-\mu(1-\frac{1}{\sqrt{2}})} & 1 
    \end{pmatrix}
\end{align}
where the diagonal elements $\langle \psi_i | \psi_i \rangle$ correspond to the squared norms of the vectors. 
Since coherent states are normalized, these are all equal to 1. 
The components $\rho_{ij}$ can be evaluated by
recalling the general scalar product between two single-mode coherent states is
\begin{equation}
    \langle \alpha|\beta \rangle = \exp\left(\alpha^*\beta - \frac{|\alpha|^2}{2} - \frac{|\beta|^2}{2}\right) 
\end{equation}
Therefore:
\begin{equation}
    \begin{aligned}
    \braket{\psi_1}{\psi_2} 
    &= \braket{\sqrt{\mu}}{0}_H\braket{0}{\sqrt{\mu}}_V
    = e^{-\mu} 
    \\
    \braket{\psi_1}{\psi_3} 
    &= \braket{\sqrt{\mu}}{\sqrt{\frac{\mu}{2}}}_H\braket{0}{\sqrt{\frac{\mu}{2}}}_V
    = e^{-\mu(1-\frac{1}{\sqrt{2}})} 
    \\
    \braket{\psi_2}{\psi_3} 
    &= \braket{0}{\sqrt{\frac{\mu}{2}}}_H\braket{\sqrt{\mu}}{\sqrt{\frac{\mu}{2}}}_V
    = e^{-\mu(1-\frac{1}{\sqrt{2}})} 
    \end{aligned}
\end{equation}
and the other terms are evaluated by using the relation $\braket{\psi_j}{\psi_i}=
\braket{\psi_i}{\psi_j}^*$.

Calculating the eigenvalues of the matrix $\rho$, we obtain:
\begin{align}
\lambda_1(\mu)&= \frac13 -\frac{e^{-\mu}}3 \\
\lambda_{2,3}(\mu) &= \frac13+\frac{e^{-\mu}}6\left(1 \pm \sqrt{1+8e^{\sqrt2 \mu}}\right)
\end{align}

Using the eigenvalues just found, we can calculate the Shannon entropy of the ensemble state $\rho$ as:
\begin{align}
    H(\mu):=-\sum_{i=1}^3 \lambda_i(\mu)\log_2(\lambda_i(\mu))
\end{align}

\section*{Acknowledgements}
\label{sec:ack}
A.D.T. acknowledges the financial support of Concessioni Autostradali Venete (CAV) S.p.A. in the framework of the doctoral scholarship agreement 38° Ciclo between CAV and the University of Padova.

This project has received funding from the European Union’s Horizon Europe Research and Innovation Programme under the project ``Quantum Secure Networks Partnership" (QSNP, G.A. No 101114043). Views and opinions expressed are however those of the author(s) only and do not necessarily reflect those of the European Union or European Commission-EU. Neither the European Union nor the granting authority can be held responsible for them.

The University of Padova is participating in the EC Funded project Nostradamus, TOPIC ID: CNECT/2023/OP/0032, in the role of contractor. It  is the goal of Nostradamus to describe the blueprint for a Testing \& Validation  Infrastructure to enable the evaluation and certification of QKD devices  and related technologies, as well as to implement and operate a  prototypical testbed facility to offer initial evaluation services which are mandatory for the accreditation from a European  security authority. The authors like to thank the whole project team for  the support and valuable exchange. Views and opinions expressed are  those of the author(s) only and do not necessarily reflect those of the European Union or the European Commission. Neither  the European Union nor the granting authority can be held responsible  for them.

The authors would like also to thank the project Q-SecGround Space of the Italian Space Agency (ASI) for providing the preliminary research basis for this work. 

\section*{Author contributions}
A.C.A., A.D.T., C.A. and D.G.M. contributed to the experimental design, tests and analyses. 
A.D.T., C.A. and G.V. contributed in the development of the theoretical models of the attack. 
The manuscript was drafted and written by A.C.A. and A.D.T., and reviewed by C.A., G.V. and P.V..

\bibliographystyle{ieeetr}
\bibliography{references}

\end{document}